# Adding high time resolution to charge-state-specific ion energy measurements for pulsed copper vacuum arc plasmas

Koichi Tanaka[1,2], Liang Han[2,3], Xue Zhou[2,4], André Anders[2]

[1] Central Research Institute, Mitsubishi Materials Corporation, 1002-14 Mukohyama, Naka-shi, Ibaraki 311-0102, Japan

[2] Lawrence Berkeley National Laboratory, 1 Cyclotron Road, Berkeley, California 94720, USA

[3] School of Physics and Optoelectronic Engineering, Xidian University, 2 South Taibai Road, Xi'an, Shaanxi 710071, China

[4] School of Electrical Engineering and Automation, Harbin Institute of Technology, Harbin, Heilongjiang 150001, China

**Abstract**     Charge-state-resolved ion energy-time-distributions of pulsed Cu arc plasma were obtained by using direct (time dependent) acquisition of the ion detection signal from a commercial ion mass-per-charge and energy-per-charge analyzer. We find a shift of energies of $Cu^{2+}$, $Cu^{3+}$ and $Cu^{4+}$ ions to lower values during the first few hundred microseconds after arc ignition, which is evidence for particle collisions in the plasma. The generation of $Cu^{1+}$ ions in the later part of the pulse, measured by the increase of $Cu^{1+}$ signal intensity and an associated slight reduction of the mean charge state point to charge exchange reactions between ions and neutrals. At the very beginning of the pulse, when the plasma expands into vacuum and the plasma potential strongly fluctuates, ions with much higher energy (over 200 eV) were observed. Early in the pulse, the ion energies observed are approximately proportional to the ion charge state, and we conclude that the acceleration mechanism is primarily based on acceleration in an electric field. This field is directed away from the cathode, indicative for a potential hump. Measurements by a floating probe suggest that potential structures travel and ions moving in the traveling field can gain high energies up to a few hundred electron-volt. Later in the pulse, the approximate proportionality is lost, which is either related to increased smearing out of different energies due to collisions with neutrals, and/or a change of the acceleration character from electrostatic to "gas-dynamic", i.e., dominated by pressure gradient.

I.  Introduction

One major feature of vacuum arc plasma is the existence of ions with high kinetic energy such as over 200 eV [1]. The mechanism of ion acceleration, however, has been in dispute for a long time. In 1960s, two different acceleration mechanisms were proposed, the "potential hump" theory [2] and the "gas-dynamic" theory [3]. Miller [4] presented a comparison between these two models with experimental data of ion velocities of $Cu^+$, $Cu^{2+}$, $Cu^{3+}$, $Bi^+$, $Bi^{2+}$ and $Bi^{3+}$. He concluded that both models roughly agree with experimental data but none in a fully satisfying way [4]. In 2000s, lots of experimental data of ion velocities measured by a time-of-flight technique were reported. With these data, Anders *et al*. experimentally established a "cohesive energy rule" which relates arc voltage, power input (for a given current), charge states, and kinetic energy of ions with the cohesive energy of the cathode material [5][6]. Yushkov *et al*. [6] expanded Miller's arguments taking into account ion acceleration in the sheath between plasma and the measuring instrument. They concluded that the velocities of each type of ion in the plasma are independent of the charge state and therefor ion acceleration is based on a gas-dynamic mechanism. In 2006, Rosen *et al*. published a comparison between two different ion velocity measurement techniques, electrostatic and time-of-flight, and concluded that the conflict between them originates from an interpretation of energy and velocity, in which peak value and averaged value are not carefully distinguished [7]. In 2013, measurements of ion energy distribution functions (IEDFs) and ion velocities of arc plasma from composite cathodes such as TiC, TiSi and TiAl reported by Zhirkov *et al*. [8][9] revealed that the peak velocities of different elements and different charge states are approximately equal. They therefore concluded that their measurements can be explained by a gas-dynamic theory

Looking at electron emission and plasma generation mechanisms at a cathode spot, Mesyats suggested a fundamental cathode mechanism in which the strong electron emission from the cathode surface leads to the generation of dense plasmas via microscopic explosions [10]. He proposed a cyclic explosion phenomenon involving heating, emission of electrons and cooling down of a micro-volume, called an "ecton", i.e. the ecton being the "quantum" of an explosive emission process. Upon expansion of the exploding micro-volume, the rate of collisions is greatly reduced. Considering the ratio of characteristic times or lengths of ionization reactions and plasma expansion one can find that changes of the plasma composition by ionization and recombination become insignificant and therefore the ratios of ion charge states become "frozen" when the plasma expands beyond "freezing region" of about 10 ~ 100 μm from the cathode spot [11] [12]. Through numerical calculations, Barengolts *et al*. showed that the increase in mean ion charge state stops at a distance of only about 1 μm from the cathode spot [13]. In 2013, in another numerical calculation, Beilis *et al*. found that the plasma parameters such as electron temperature and degree of ionization are high at an initial stage of cathode spot development, and then decrease to reach steady-state after about 1 μs [14].

Kinetic energies of ions and their energy distributions are, however, affected by particle collisions. Generally, a fraction of the momentum of a particle is transferred to another particle upon collision, and the kinetic energy is thermalized, i.e. shared among all particles. Charge exchange collisions play an especially important role as the cross section of such collisions is large. A reduction of high ion velocity and mean charge state of pulsed arc plasma as a function of time was reported by Anders *et al*., who stressed the evolution of neutral particle density and related onset of charge transfer processes in the first few 100 μs after arc ignition [5][15][16]. At the very beginning of discharge, when the neutral particle density is still low, ions collide less and can travel though vacuum adjacent to the cathode and approximately maintain their velocities reached near the cathode spot. Later into the pulse, the space is filled with plasma and neutrals from previous explosive mission processes, the neutral density is increased, and ions are more likely to collide with neutrals and thereby reduce their average velocity in the direction of the plasma flow [17]. Hence, IEDFs change as a function of time, and the discussion of the ion acceleration mechanism needs to also include the effects of particle collisions and changes outside the immediate vicinity of cathode spots.

Most of the numerous studies are based on time-integrated measurements, and less research deals with time-resolved measurement of ion energies. One possible way to achieve time resolution is to use the spectrometer's gating feature, which allows data acquisition for a certain time window (the "gate") by suitable electric signal controls [18][19][20][21]. This method, however, cannot be used for high resolution measurement like under 10 μs because the ion signal intensity is reduced as the gate window narrows. An additional complication in terms of a high time resolution IEDF measurements lays in the fact that the time of flight differs for each mass/charge [22]. Karkari and Voronin developed and improved the technique of ion energy distribution measurements with a high time-resolution of 1 μs employing a gating grid at the instrument's entrance orifice [22][23]. In pulse discharges, furthermore, IEDFs are affected by plasma potential transient. Bradley *et al.* combined IEDF measurements and time-resolved potential measurements of bi-polar magnetron discharge showing that IEDFs have three peaks correlated to distinct phases of discharge voltage which bring distinct plasma potentials [24].

In this paper, in order to gain deeper understanding of the plasma evolution in a cathodic arc discharge, charge-state-resolved ion energy-time-distribution function (IETDF) measurements were carried out with a time resolution better than 1 µs. In order to obtain this high time-resolution, the raw ion detection signal from a commercial mass and energy analyzer was directly acquired and processed on an broad-band digital oscilloscope. In the following we explain the principle and discuss our findings.

## II. Experimental Section

**General setup**

The general experimental set up is shown in Fig. 1. The experiment was done with a miniature-gun arc plasma source and an integrated quadrupole mass and energy analyzer (HIDEN EQP300) mounted to a cylindrical vacuum chamber of 1 m inner diameter and 0.25 m inner height. The chamber was pumped by an oil-free mechanical pump and a cryogenic pump to a base vacuum pressure in the $10^{-4}$ Pa region. Principle and a schematic diagram of the miniature-gun arc plasma source was described by MacGill *et al*. [25]. An oxygen-free Cu (99.94%) rod of 6.35 mm (1/4 inch) diameter and about 20 mm length was used as a cathode. The cathode was set in front of the entrance orifice of EQP300 and directed to it. Two distances between cathode and orifice were investigated: 0.18 m and 0.38 m.  A pulse forming network (PFN) [26] with 9 LC-sections and one RC-section served as power supply delivering a current pulse of approximate rectangular shape. A silicon controlled rectifier (SCR) was put between the output leads to make a short circuit in order to rapidly terminate the discharge. The trigger signals for arc ignition and termination were controlled by combining two pulse delay generators (TGP110 by TENMA and MODEL-214A by Hewlett Packard) to obtain a 600 μs long rectangular current pulse as shown Fig. 2. The capacitors in the PFN were charged by a DC power supply (Series-KL by Glassman High Voltage Inc., maximum 1 kV and 3 A) and discharge current was monitored on an oscilloscope (TDS5054B-NV by Tektronix) by a wide-band current transformer (101X by Pearson Electronics Inc., 0.01 VA$^{-1}$). The cathode voltage (relative to the grounded anode) was measured by a 1:100 high voltage probe (P5100A by Tektronix). The flat portion of current pulse was 190 ~ 200 A with the charging voltage of 381 V. The repetition rate of the discharge was kept constant at 3 pulses per second for all measurements.

**Time-resolved Ion Energy Measurements**

Generally, the energy distributions of ions of various charge states can be measured by combining an electrostatic energy filter and a quadrupole mass analyzer. Strictly speaking, with this kind of equipment, the ions are filtered by $E/Q$ and $m/Q$, where $E$, $m$ and $Q$ are kinetic energy, mass and charge state of ions, respectively. In contrast to conventional data acquisition operated by the manufacturer's software, the raw pulse signals from the detector unit were acquired in our experiments from the EQP control unit via its auxiliary output port. Raw ion detection pulse signals are negative with respect to a reference offset of 4.8 V (Fig. 3). Each individual detection pulse has a width of 10 ns, however, due to reflections caused by impedance mismatch at the oscilloscope, the actual detection pulse is distorted and several 100 ns wide. This pulse shape distortion does not affect the measuring principle as each detection pulse has the *same* shape and amplitude. Let us designate $n_{on}(t)$ as the number of detection pulses at a given time $t$ after ignition for $N$ discharge pulses, with $n_{on}(t) \leq N$. The voltage $\overline{V}(t)$ recorded by the oscilloscope, averaged over $N$ pulses, can be calculated as

$$\overline{V}(t) = \frac{(N - n_{on}(t))V_{off} + n_{on}(t)V_{on}}{N} \tag{1}$$

where $V_{off}$ is the OFF-state voltage and $V_{on}$ is the ON-state voltage. The same expression can be used to explicitly relate $n_{on}(t)$ to the measured voltage:

$$n_{on} = \frac{(\overline{V}(t) - V_{off})}{V_{on} - V_{off}} N. \tag{2}$$

We can define the function $I(t)$ as the "intensity" of the ion detection signal:

$$I(t) \equiv -(\overline{V}(t) - V_{off}) \propto n_{on}(t). \tag{3}$$

We note that $I(t)$ has the minus sign in Eq. (3) since individual ion detection pulses are negative with respect to the reference potential $V_{off}$. In this study, $N$ was set to 50 (the average setting at the oscilloscope).

In the practical measurements, the detector signal also contains low frequency noise induced by the high discharge current. Additionally, an actual pulse signal seen on an oscilloscope may have high frequency ringing in the ~50 MHz range. However, since the shape of the individual detector pulses is always the same, deviations from the perfect rectangular TTL signal do not have an influence on the waveform of $I(t)$ on a timescale of 100 μs. We note that (i), the signal $I(t)$ may saturate and therefore, to avoid saturation, the electron multiplier voltage was selected to be a relatively low value, and (ii), the actual background potential $V_{off}$ was used to account for low frequency noise. An example signal waveform $I(t)$ is shown in Fig. 3. $I(t)$ can be measured at various electrical settings of EQP300 system in which detected ions are limited to certain *E/Q* and *m/Q* ratios, therefore $I(t)$ is a function of not only time *t* but also kinetic energy *E*, mass *m* and charge state *Q*, i.e., we have $I(t, E, m, Q)$, which we called the ion energy-time-distribution functions (IEDTFs).

Here we recall the general principle of the EQP300 instrument to describe a proper treatment of time-of-flight effects. The kinetic energy $E_{kin}$ of an ion inside the instrument can be described as

$$E_{kin} = E_0 - Qe(V_{local} - V_{plasma}) \qquad (4)$$

where $E_0$ is kinetic energy of ions in the plasma, *e* is elementary charge, *Q* is charge state number, $V_{local}$ is the local potential inside the instrument, and $V_{plasma}$ is the plasma potential of the location outside the instrument, from which the ion entered the sheath of the entrance aperture. $V_{local}$ varies at each section of the instrument like

$$V_{local} = V_{reference} - V_{axis}, \qquad (5)$$

where $V_{axis}$ is the potential of the axial section after the entrance relative to the instrument's reference potential (for even more detail on the potentials see the instrument's user manual). In order to pass the energy filter (E-sector) and to proceed to the quadrupole mass sector (m-sector), an ion is required to have the kinetic energy $Q \times E_{filter}$, where $E_{filter}$ is an energy value determined by a geometric factor and the voltage difference of plates in the electrostatic energy filter. Therefore, to pass the E-sector, the kinetic energy of the ion must satisfy the relation

$$E_{kin}^{E\text{-sec}} = E_0 - Qe(V_{reference} - V_{axis} - V_{plasma}) = QE_{filter}. \qquad (6)$$

In our experiment, by setting the value of $E_{\text{filter}}$ and $V_{\text{axis}}$ to 40 eV and 40 V, respectively, the kinetic energy of an ion which is detected at the detector can be simply calculated as

$$E_0 = Qe\left(V_{\text{reference}} - V_{\text{plasma}}\right). \tag{7}$$

By varying the value of $V_{\text{reference}}$ from 6/$Q$ V with an increment step of 6/$Q$ V for each $m/Q$ of 63.00 (Cu$^{1+}$, for amu 63, the most abundant isotope of copper), 31.50 (Cu$^{2+}$), 21.00 (Cu$^{3+}$) and 15.76 (Cu$^{4+}$), IETDFs of ions with different charge states were obtained. The electron multiplier voltage was set to 1.9 kV for all measurements, low enough to avoid intensity saturation.

Compared to the discharge duration of a few 100 μs, the time of flight $\tau_{\text{TOF}}$ of an ion is 10 μs ~ 150 μs, which is not negligible. Therefore, the intensity $I(t)$ should be time-corrected by

$$t = t_{\text{raw}} - \tau_{\text{TOF}} \tag{8}$$

where $t_{\text{raw}}$ is the time of the raw waveform recorded by the oscilloscope after ignition. The total time-of-flight $\tau_{\text{TOF}}$ can be described as sum of the flight times in each sector:

$$\tau_{\text{TOF}} = \tau_{\text{plasma}} + \tau_{\text{ext}} + \tau_{\text{E-sec}} + \tau_{\text{m-sec}} + \tau_{\text{det}} \tag{9}$$

where $\tau_{\text{plasma}}$, $\tau_{\text{ext}}$, $\tau_{\text{E-sec}}$, $\tau_{\text{m-sec}}$ and $\tau_{\text{det}}$ are time of flight of an ion in the plasma region, extractor, E-sector, m-sector and detector region, respectively. Each time of flight is calculated according to Eq. (10) to Eq. (14):

$$\tau_{\text{plasma}} = \frac{L_{\text{plasma}}}{\sqrt{\dfrac{2E_0}{m}}} \tag{10}$$

$$\tau_{\text{ext}} = \frac{L_{\text{ext}}}{\sqrt{\dfrac{Qe|V_{\text{ext}}|}{2m}} + \sqrt{\dfrac{Qe|V_{\text{axis}}|}{2m}}} \tag{11}$$

$$\tau_{\text{E-sec}} = \frac{L_{\text{E-sec}}}{\sqrt{\frac{2E_{\text{kin}}^{\text{E-sec}}}{m}}} \tag{12}$$

$$\tau_{\text{m-sec}} = \frac{L_{\text{m-sec}}}{\sqrt{\frac{2E_{\text{kin}}^{\text{m-sec}}}{m}}} \tag{13}$$

$$\tau_{\text{det}} = \frac{L_{\text{det}}}{\sqrt{\frac{Qe|V_{\text{dynode}}|}{2m}}} \tag{14}$$

where $L_{\text{plasma}}$, $L_{\text{ext}}$, $L_{\text{E-sec}}$, $L_{\text{m-sec}}$, $L_{\text{det}}$, are the length of the plasma region (i.e. distance between cathode surface and entrance orifice of the instrument), extractor, E-sector including axial section before the energy filter, m-sector, and detector, respectively; $V_{\text{ext}}$ and $V_{\text{dynode}}$ are the potentials of extractor and dynode with respect to $V_{\text{reference}}$, and $m$ is the mass of an ion (63 amu, $1.05 \times 10^{-25}$ kg for Cu). The ion energy in the m-sector can be calculated as

$$E_{\text{kin}}^{\text{m-sec}} = E_{\text{kin}}^{\text{E-sec}} - Qe(V_{\text{axis}} - V_{\text{transmission}}) = QeV_{\text{transmission}}. \tag{15}$$

In our experiment, $V_{\text{transmission}}$ was set to 3 V, therefore $E_{\text{kin}}^{\text{m-sec}}$ is simply $Q \times 3$ eV. In this work, $L_{\text{ext}}$, $L_{\text{E-sec}}$, $L_{\text{m-sec}}$, $L_{\text{det}}$, $V_{\text{ext}}$ and $V_{\text{dynode}}$ were 0.059 m, 0.265 m, 0.179 m, 0.041 m, -12 V and -1200 V. $\tau_{\text{ext}}$, $\tau_{\text{E-sec}}$, $\tau_{\text{m-sec}}$ and $\tau_{\text{det}}$ are independent of the kinetic energies of ions.

**Plasma potential and floating potential measurements**

In order to determine the kinetic energy of ions in plasma, the plasma potential was measured by an emissive probe. A tungsten filament with the diameter of 0.075 mm was bent into a 2 mm diameter semicircle loop and mounted on insulating alumina tube with 2 bores and touched by Cu wires inside the bores similar to a probe described in Fig. 1 of Ref. [27]. The probe was set in front of the cathode at a distance of 21 mm. The filament was connected to a DC power supply (72-7295 by TENMA, max 40 V and 5 A) through a switching unit as shown in Fig. 4. The potential of the filament was measured by an oscilloscope triggered by the rising edge of arc pulse. The switching unit between the filament and the heating power supply is to isolate the filament from the heating circuit while a gate signal is applied to p- and n-type MOSFETs (FQP17P10 and IRF740). The isolation resistivity of the MOSFETs was over 2 MΩ, which is higher than the internal resistivity of the oscilloscope, 1 MΩ. The potential of the filament during measurements can therefore be regarded as the floating potential. MOSFETs in the switching circuit were set to isolation mode 50 μs before the arc ignition and kept there for 1 ms. As e.g. Rauch *et al*. [27] described, the relationship between plasma and floating potentials is given by,

$$V_{\text{plasma}} = V_{\text{floating}} + \frac{kT_e}{e\overline{Q}} \ln\left(\frac{I_{\text{es}}}{I_{\text{is}} + I_{\text{emission}}}\right) \quad (16)$$

where $V_{\text{plasma}}$ is a plasma potential, $V_{\text{floating}}$ is floating potential, $k$ is Boltzmann constant, $T_e$ is electron temperature, $e$ is elementary charge, $\overline{Q}$ is mean charge number of ions arriving at the probe, $I_{\text{es}}$ is the electron saturation current from plasma, $I_{\text{is}}$ is the ion saturation current from plasma and $I_{\text{emission}}$ is the thermally emitted electron current from a filament. As long as thermal emission of electrons from the filament is significant, the second term approaches zero and the floating potential is approximately equal to the plasma potential. The probe potential was averaged for 20 pulses on an oscilloscope. The probe potential during the arc pulse increased from -5 V to -1.5 V as the heating current increased from 0 A (i.e., cold probe) to 1.3 A and remained constant between 1.4 A and 1.6 A (hot, emissive probe).

To see the delay of floating potential fluctuation, two Cu wires with diameter of 0.6 mm ea. covered by ceramics tube except 2 mm from the top were set at 11 mm and 41 mm in front of the cathode. The floating potentials of these probes were measured by an oscilloscope.

**III.     Results**

The plasma potential measured by an emissive probe drops rapidly after ignition and then increases after 4 μs as shown Fig. 5. After about ~ 50 μs, the plasma potential remains approximately constant for the rest of the discharge pulse. The potential value of -1.5 V with the heating current of 1.6 A was used for kinetic energy and time of flight calculations. The deepest point of floating potentials measured at different positions, 11 mm and 41 mm from the cathode surface, showed a delay time of approximately 2 μs as shown Fig. 6.

Time evolution of ion detection intensities were measured for $Cu^{1+}$, $Cu^{2+}$, $Cu^{3+}$ and $Cu^{4+}$, with the reference potential in the range from 6 eV up to 408 eV. No higher charge states than $Cu^{4+}$ were observed. Time evolution of the ion detection intensities at 0.18 m and 0.38 m from the cathode are shown in Figs. 7 (a) to (d) and Figs. 8 (a) to (d), respectively. In both cases, ions with energies greater than 100 eV appear first, and then the intensity of lower energy ions increases for a long duration like 300 μs as can be seen in the results of $Cu^{2+}$, $Cu^{3+}$ and $Cu^{4+}$.   It appears that ion energy distributions shifted to lower energy. One can see that the energy of multiply charged ions at 50 μs vary and higher charge state ion shows higher energy range.    In another words, ion energy of multiply charged ion are reduced as time elapses. Unlike other charge states, the intensity of $Cu^{1+}$ increased when the distance was increased from 0.18 m to 0.38 m from the cathode.

Another important finding in Figs. 7 and 8 is the existence of ions with high energy, some ions even exceeding 150 eV in the first 50 μs after ignition. In the first 50 μs, the kinetic energy range depends on the charge ion state, e.g. approximately up to 200 eV for 2+, 290 eV for +3 and 410 eV for 4+.

To see the evolution of total intensities of each ion species ($m$ = 63 amu, $Q$) and time dependence of their average, $\overline{Q}(t)$, the intensities at a given time were summed up over all energies,

$$G_{Q+}(t) = \frac{1}{Q} \sum_{\text{all } E} I(t, E, Q, m), \tag{17}$$

and

$$\overline{Q}(t) = \frac{\sum_{Q=1}^{4}\left(Q \cdot G_{Q+}(t)\right)}{\sum_{Q=1}^{4} G_{Q+}(t)}. \tag{18}$$

Because of a deep plasma potential dip at the beginning of the discharge pulse; the $\tau_{\text{plasma}}$ calculation is not applicable for the first 50 μs. Due to the negative potential, ions are rather decelerated in front of the instrument's grounded orifice. The kinetic energy is a function of time,

$$E_0(t) = Qe\left(V_{\text{reference}} - V_{\text{plasma}}(t)\right). \tag{19}$$

The actual kinetic energy of ions in the plasma in the beginning of the discharge is much higher than what could be determined with $V_{\text{plasma}}$ of -1.5 V, which is the value during the plateau phase of the discharge. Ions in the beginning have higher velocities and their real $\tau_{\text{plasma}}$ is much shorter than later in the pulse. Therefore, the regions before 50 μs and after 500 μs are excluded from the results. For both of distances of 0.18 m and 0.38 m, $\overline{Q}(t)$ showed a rapid decrease in the time period $t$ = 50 ~ 150 μs after ignition, as shown in Figs. 9 (a) and (b). Comparing the decay in 200 μs ~ 500 μs after ignition, one can find that the intensity of $Cu^{1+}$ increases, despite the fact that the discharge current is almost constant, and $Cu^{2+}$, $Cu^{3+}$ and $Cu^{4+}$ decrease slightly at a distance of 0.38 m, while they are approximately constant at 0.18 m.

## IV. Discussion

Here we divide the plasma pulse into 2 stages, which are dramatically different in plasma potential: (A) the initial plasma expansion phase (0 - 50 μs) and (B) the steady-state phase (50 - 600 μs).

(A) The initial plasma expansion phase

In this phase, the cathode spot plasma expands into vacuum. The charge state dependence of the ion energies suggests the dominance of an electrostatic ion acceleration mechanism. Since we can detect ions of high energy even when the plasma potential drops rapidly to -30 V and recovers to -1.5 V, very high energy ions are present in this phase. As the plasma potential is negative with the respected to the grounded entrance orifice potential, the kinetic energy of each ion species is even higher than the kinetic energy of ions calculated by Eq. (7).

The existence of such high energy ions can be associated with traveling potential structures as recently proposed [28]. The emissive probe and Langmuir probe measurements shown in Figs. 5 and 6 support the concept that the potential's spatial distribution is not uniform or steady when the initial plasma is expanding. One can suppose from these results that "traveling potential fields" exist and (i) the potentials at various point of the space become negative due to electron expansion which is much faster than that of ions; (ii) after the negative potential is formed, ions generated at the cathode spot accelerate into that space due to the local potential gradient (electric field), the potential increases and approaches the steady-state plasma potential, (iii) ions near the deep potential well are more accelerated and gain higher kinetic energy than following ones, when the potential dip has become shallower. Compared to the velocities of ions obtained, such as $0.4\times10^4$ m/s for 6 eV and $3.0\times10^4$ m/s for 300 eV, the velocity of the negative potential peak such as $1.5\times10^4$ m/s is enough to support this acceleration mechanism. One can find this interpretation reasonable because the initial plasma expansion is dominated by electrons that escape faster from the dense plasma than the positive ions.

(B) The constant potential phase

The existence of a decay of ion energies while the plasma potential doesn't show any fluctuations (Fig. 5) suggests that particle collisions increase for some time after arc pulse ignition. Since this time scale of the decay of ion energy ~ 100 μs is much longer than the estimated life time of cathode spot of 10 ~ 40 ns [29], the decay is mainly caused by particle collisions *relatively far from* the cathode spot. This interpretation further suggests that the density of particles the ions interact with increases even as the plasma potential remains relatively constant in this phase.  As the plasmas fills the space and the sticking coefficient of ions hitting walls and other objects is somewhat less than unity, the background neutral density increases followed by a collisional slow-down of ions coming from the cathode spot.  The energy distributions shift to the lower energy range and finally the ion energy distributions of multiply charged ions can overlap with each other, forming a merged peak as previously reported [7][16].

The increase of $Cu^{1+}$ ion in the later part of the pulse, as seen in Figs. 7(a) and 8(a), suggests that singly charged ions are generated by interactions of particles in the plasma and not in the cathode spot. A most straight-forward explanation of $Cu^{1+}$ ion generation outside the cathode spot are charge transfer collisions of the type

$$M^{n+} + M^0 \to M^{(n-1)+} + M^{1+} \quad (1)$$

as was proposed earlier [15]. Thus, this reaction needs the existence of a neutral atom and increases the $Cu^{1+}$ density. Neutral atoms can be generated by neutralization of ions (albeit unlikely, see below), or by self-sputtering and perhaps most importantly by a less-than-perfect sticking of condensable ions such as copper [17]. The filling time of the volume between cathode and the detector with neutral atoms is of the order of 100 μs, which roughly agrees with the experimental data of the decay of mean ion charge state and velocities [15][17]. The thus-obtained decay of kinetic energies of $Cu^{2+}$, $Cu^{3+}$ and $Cu^{4+}$ in this study is consistent with this physical interpretation. Furthermore, the observation that more $Cu^{1+}$ ions are detected at 0.38 m than at 0.18 m also strengthens the arguments for the charge exchange mechanism. It also provides a clue for the "length scale" of charge state and energy reductions. This means that the ion charge states and energy distributions are functions of time and distance until they reach approximate steady-state. The typical characteristic times and lengths for reaching steady-state are milliseconds and meters, respectively. As the arc pulse is made longer, and in the limit is direct current, the measured ion charge state and energy distributions reflect the steady-state phase of the experiment here. Charge exchange reactions and generation of singly charged ions influence the mean ion charge state $\overline{Q}$ as shown Fig. 9. Another possible reason for increase of singly charged ions could be the recombination of $Cu^{2+}$

$$Cu^{2+} + e^- \to Cu^{1+}. \quad (2)$$

This reaction, however, is unlikely to take place unless a third body is present which is needed to ensure energy and momentum conservation in the recombination reaction. Since any of positive ions cannot be the third body because of Coulomb repulsion, only neutrals can be the third body.

The probability of three particles collisions, e.g. ion-electron-neutral collisions, is less than the probability of two particle collision like Eq. (1) and therefore negligibly small.

It is known that cathode processes, and even direct current (dc) operation, are not continuous processes but consisted of a number of microexplosions at cathode spots [29][30][31]. Assuming a microexplosion leads to a potential hump due to electrons leaving the dense plasma faster, and potential structure moves from the cathode spot to the chamber space like a wave, the origin of acceleration mechanism of positive ions at cathode spot can be explained by the "potential hump theory" which has been dismissed for some time but was recently revived [28]. Based on the observations of the temporal evolution of ion energies in pulsed Cu arc plasma, we propose here an ion generation and acceleration model of an arc discharge as follows: (i) multiply charged ions are formed due to electron impact ionization in the dense plasma at cathode spot, (ii) high dense plasma forms high potential region ("potential hump") in cathode spot as electrons leave the dense plasma first, (iii) ions are accelerated into chamber space by electric field of the hump, (iv) a quasi-continuous flow of plasma is stabilized by ion-electron friction and charge exchange collision so that all ion species fly with essentially the same velocity: the hump acceleration mechanism transitions to a more gas-dynamic mode which is still supplemented by the effects of transient potential humps. Experimental findings indicate that the transition to a quasi-stationary flow takes time, a few hundred μs, and distance, at least 0.1 m in the Cu case. These characteristic time and distance may depend on the element and could play an important role to determine kinetic energy and charge state reduction of ions coming from the cathode surface.

## V.     Summary and Conclusions

Charge-state-resolved ion energy-time-distributions (IETDFs) of pulsed Cu arc plasma were measured by using a direct acquisition method of the ion detection signal from a commercial mass and energy analyzer. Shifts of energies of $Cu^{2+}$, $Cu^{3+}$ and $Cu^{4+}$ ions to lower values for a few 100 μs after arc ignition strongly suggest that collisions in the plasma, and specifically charge exchange collisions, are important. This is further supported by the generation of $Cu^{1+}$ ions in the later part of the pulse, an increase of the $Cu^{1+}$ intensity at a greater distance of 0.38 m from the cathode surface, and a slight decrease of the mean charge state of Cu ions observed at further distance of 0.38 m. At the very beginning of the pulse, when the plasma expands into vacuum and the plasma potential dips deeply to negative values, ion with much higher energy were observed. Because the energy range of these ions is approximately proportional to their ion charge state, we can conclude that the acceleration mechanism is based on electric field acceleration. As measured by a floating probe, the deepest point of the potential valley, formed by a rapidly expanding electron cloud, had a short delay time of 2 μs to travel from a probe at 11 mm to a probe at 41 mm. The negative potential layer caused by electrons and the quasi-neutral plasma following the moving layer accelerate ions which can gain very high energies up to a few 100 eV.

**Acknowledgment**s

We gratefully acknowledge the Mitsubishi Materials Corporation for supporting this study under Contract No. WF010678. Work at Berkeley Lab is supported by the U.S. Department of Energy under Contract No. DE-AC02-05CH11231.

**Figure Captions**

Fig. 1 Schematic diagram of experimental set up with a Pulse Forming Network (PFN) of nine LC sections and one RC section, a miniature-gun type arc plasma source and a mass and energy analyzer (EQP300). The vacuum arc discharge was ignited and terminated by SCR-1 and SCR-2 controlled by pulse delay generators.

Fig. 2 Arc current and voltage during the pulse, averaged over individual 50 discharge pulses.

Fig. 3 Background potential $V_{\text{off}}$ and example of an averaged ion detection pulse signal waveform $\overline{V}(t)$, and calculated ion detection intensity $I(t)$ ($m/Q$ = 63.0 amu ($Cu^{1+}$), $V_{\text{reference}}$ = 30 eV at 0.38 m from cathode). All signals were averaged over 50 discharge pulses.

Fig. 4 Schematic diagram of probe measurement setup.

Fig.5 Potentials of the tungsten filament probe during the pulse with and without heating, averaged over 20 discharge pulses.

Fig.6 Time evolution of the floating potential at 11 mm and 41 mm from cathode, averaged over 20 discharge pulses.

Fig. 7 Intensities of ion detection $I(t)$ of (a) $Cu^{1+}$, (d) $Cu^{2+}$, (c) $Cu^{3+}$ and (d) $Cu^{4+}$ at different kinetic energies and times at a distance of 0.18 m. Contour axis is intensity in V. The time axis is adjusted by subtracting $\tau_{\text{TOF}}$ for each curve.

Fig. 8 Intensities of ion detection $I(t)$ of (a) $Cu^{1+}$, (d) $Cu^{2+}$, (c) $Cu^{3+}$ and (d) $Cu^{4+}$ at different kinetic energies and times at a distance of 0.38 m. Contour axis is intensity in V. The time axis is adjusted by subtracting $\tau_{\text{TOF}}$ for each curve.

Fig. 9 Energy-integrated intensity $G_{Q+}(t)$ and mean charge state $\overline{Q}$ at a distance of (a) 0.18 m

and (b) 0.38 m. Time intervals 0 - 50 μs and > 500 μs are eliminated from the plots because $\tau_{TOF}$ calculation is not applicable due to plasma potential changes.

**Figures**

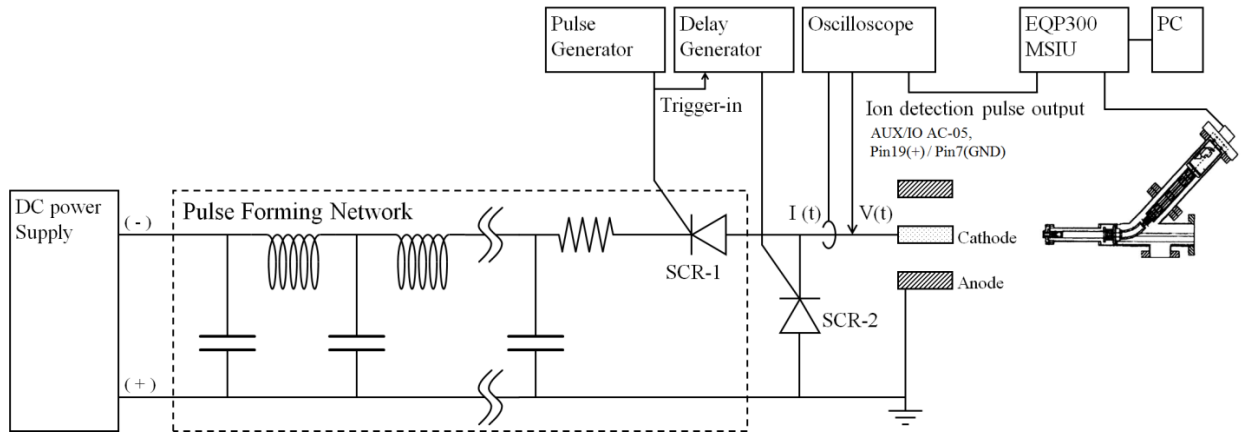

Fig. 1

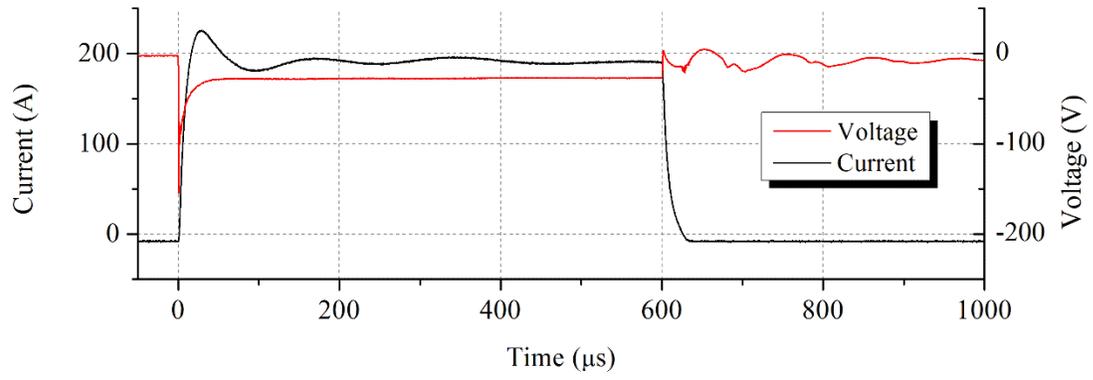

Fig. 2

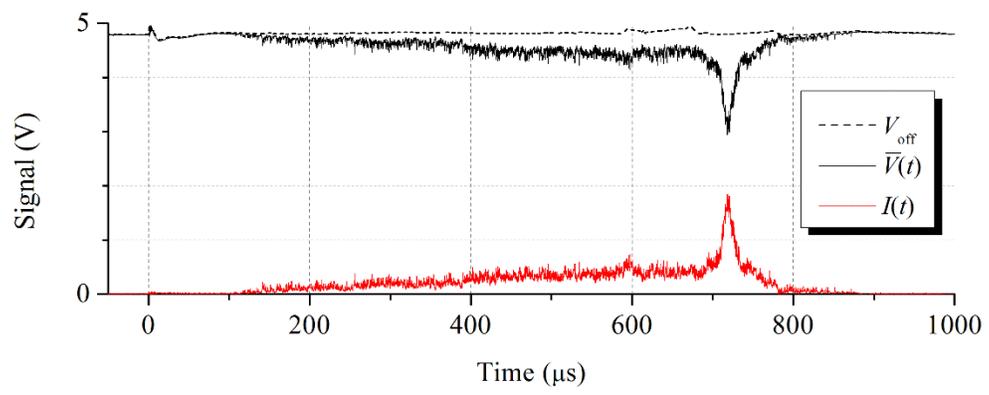

Fig. 3

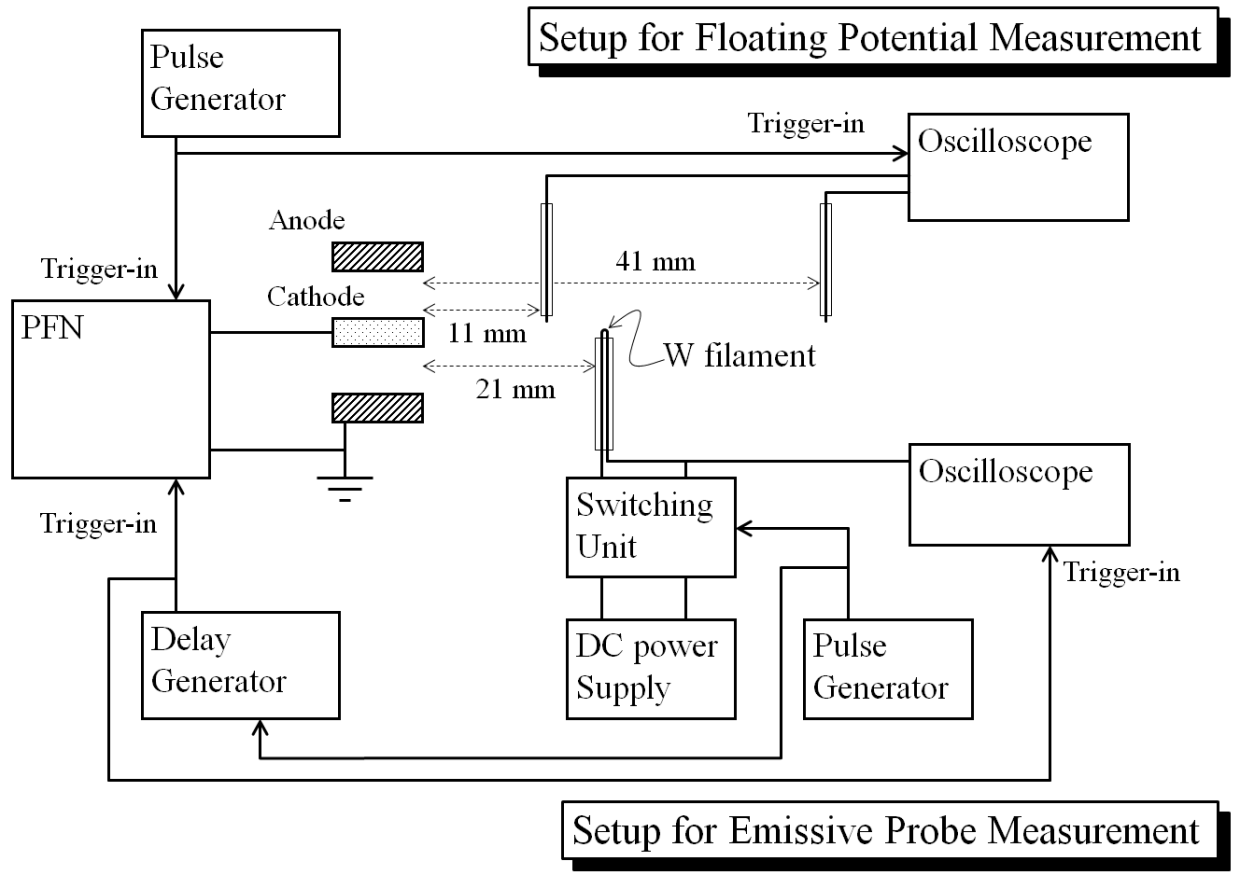

Fig. 4

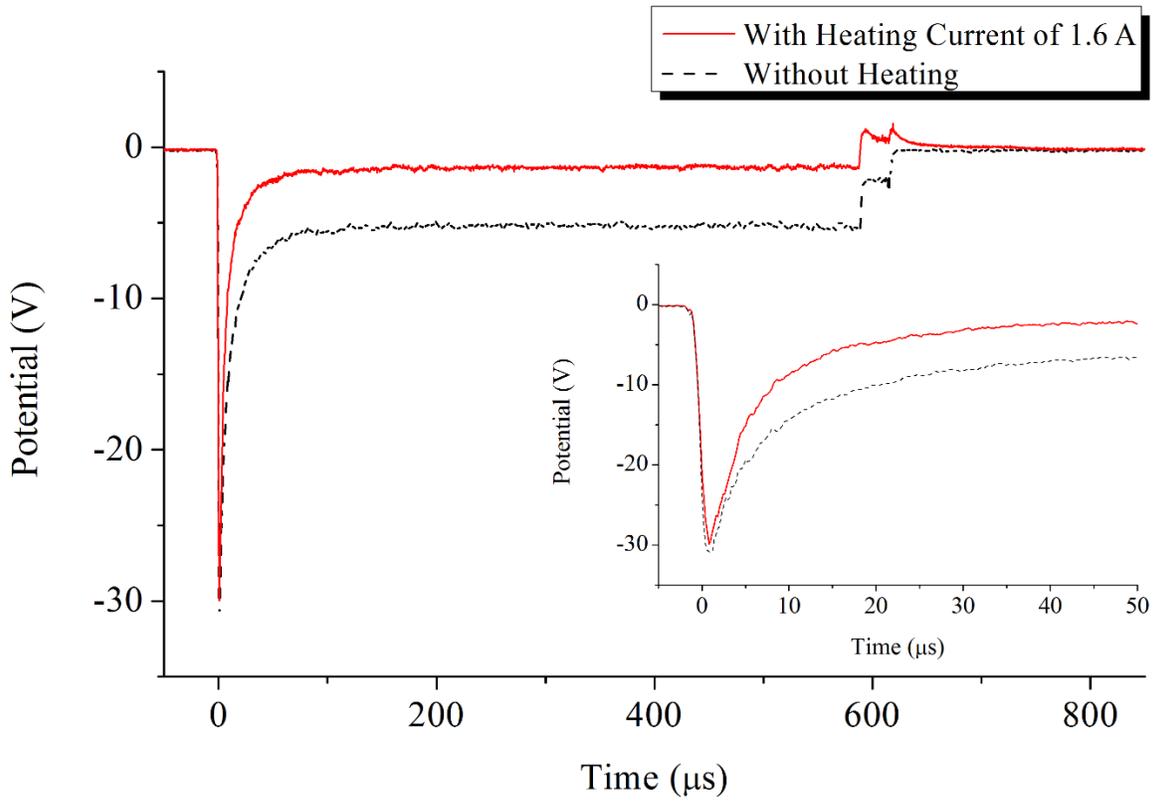

Fig. 5

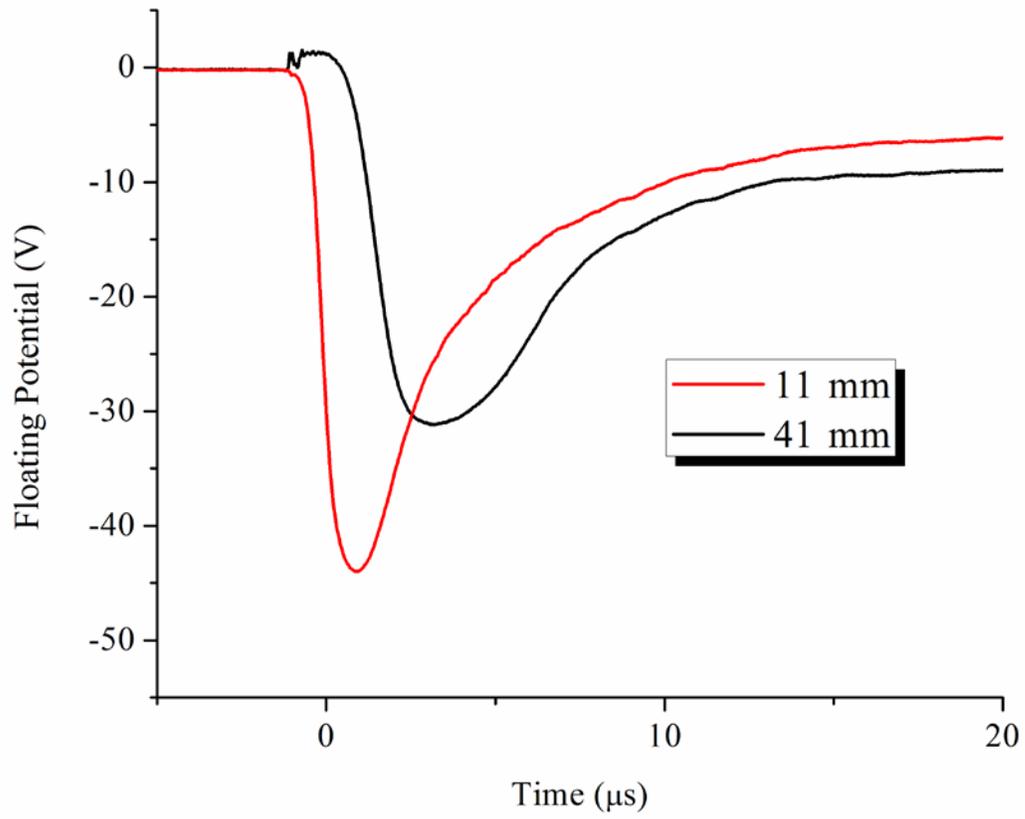

Fig. 6

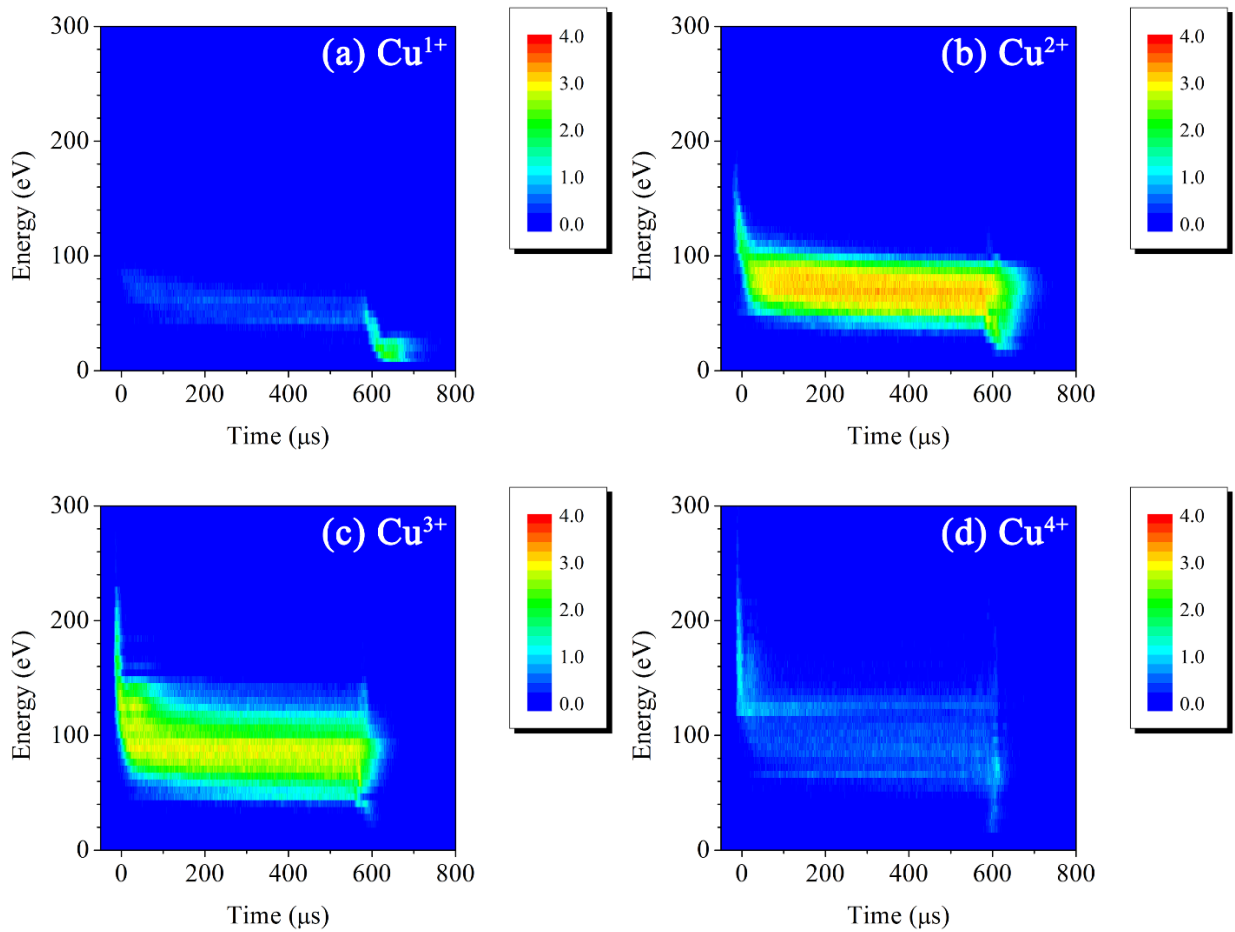

Fig. 7 (a) - (d)

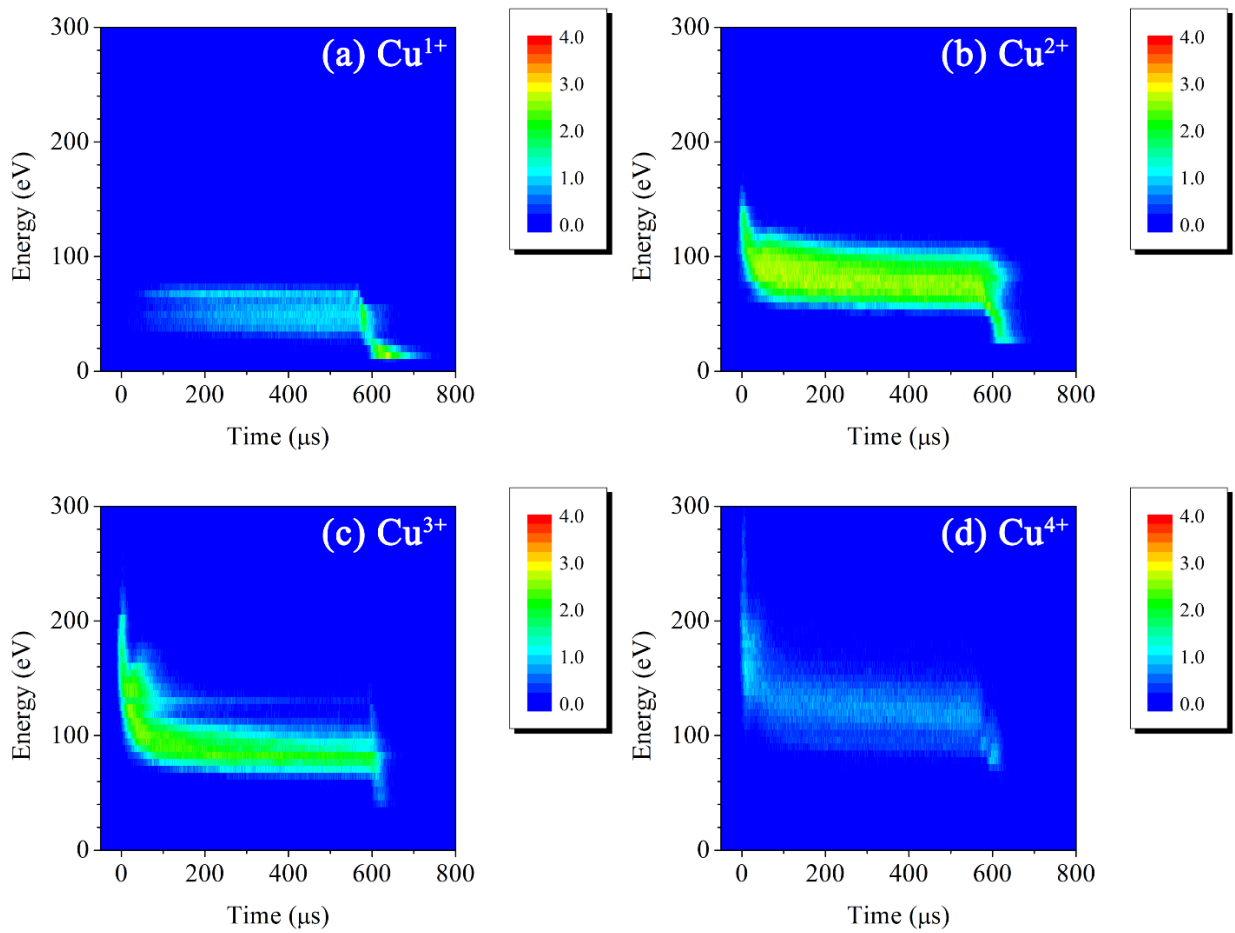

Fig. 8 (a) - (d)

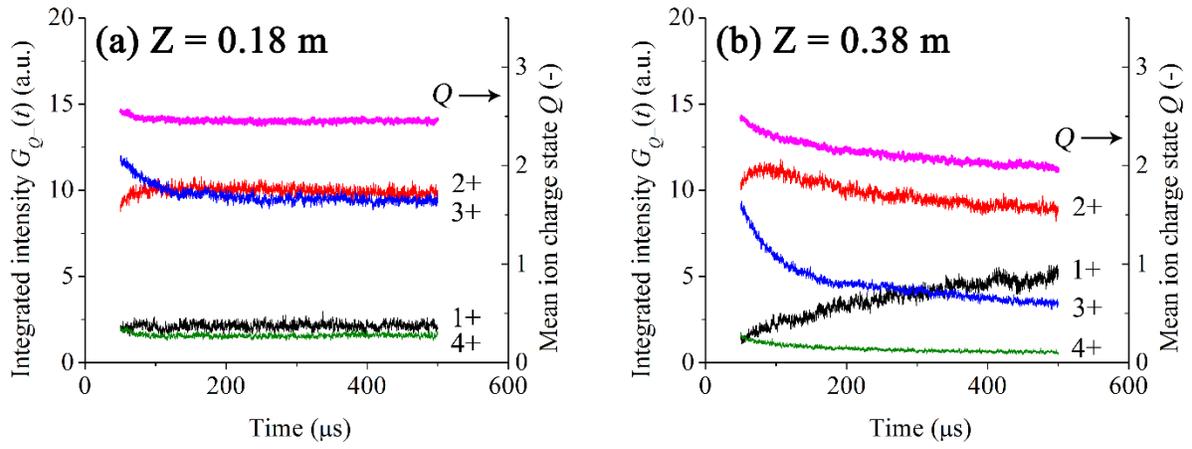

Fig. 9 (a) - (b)